\documentclass[aps,preprint,groupedaddress,showpacs]{revtex4}

\usepackage{graphicx}
\usepackage{amsmath}
\usepackage{amssymb}
\usepackage{hyperref}
\usepackage{color}

\begin{document}

\title{On the statistics of edge fluctuations: comparative study between various fusion devices}

\author{F. Sattin, M. Agostini, P. Scarin, N. Vianello, R. Cavazzana, L. Marrelli, G. Serianni}

\affiliation{Consorzio RFX, Associazione EURATOM-ENEA sulla fusione, Corso Stati Uniti 4,
Padova, Italy}

\author{ S.J. Zweben}

\affiliation{Princeton Plasma Physics Laboratory, Princeton, NJ 08543, USA}

\author{R.J. Maqueda}

\affiliation{Nova Photonics, NJ 08540, USA}

\author{Y. Yagi, H. Sakakita, H. Koguchi, S. Kiyama, Y. Hirano}

\affiliation{National Institute of Advance Industrial Science and Technology (AIST),
Tsukuba, Ibaraki, 305-8568, Japan}

\author{J.L. Terry}
\affiliation{MIT Plasma Science and Fusion Center, Cambridge, MA 02139, USA}

\begin{abstract} 
In this paper we present a statistical study of edge fluctuations taken with the Gas Puffing Imaging 
(GPI) diagnostics. We carry out a comparison of GPI signal from an extensive 
database including four devices (two Tokamaks and two Reversed Field Pinches).
The data are analyzed in terms of their statistical moments Skewness and Kurtosis, as done in 
[B. Labit, et al, Phys. Rev. Lett. \textbf{98}, 255002 (2007)]. The data align along parabolic curves, although different from machine to machine, with some spread around the best-fitting curve.
A discussion about the meaning of the parabolic trend as well as the departure of real data from it is provided.  
A phenomenological model is finally provided, attempting to accomodate experimental evidence.
 
\end{abstract}

\pacs{52.35.Ra, 52.25.Xz, 52.35.Kt, 52.55.Hc}

\maketitle

\section{Introduction}
Edge plasma turbulence represents a main obstacle towards a good confinement of magnetic fusion plasmas,
and has been therefore studied for several years. Its understanding has considerably progressed,
and nowadays the results produced by first-principle drift-interchange turbulence numerical models 
share several statistical features with experimental ones \cite{modelli}. However, we have not yet grasped
 a complete comprehension of this problem. \\
One feature of edge turbulence phenomenology that has arisen interest of researchers is its (supposed) 
universality: statistical properties of edge turbulence appear quite similar regardless of the device 
they are measured in. Indeed, if the driving of turbulence is everywhere provided by the same instabilities, 
it appears quite natural that some affinities must appear. Attempts of applying principles of universality 
to edge plasmas may be traced back since the works on Self-Organized-Criticality ideas  \cite{soc}. It was however soon realized that, if one goes beyond the qualitative similarity of the signal, 
at the quantitative level relatively large differences apparently exist between different machines  (see, e.g. the lack of consensus about the analytical form of the PDF of fluctuations) \cite{pdf,pop04,ppcf06,labit,mast,pedrosa}. 
Differences are unavoidable, since the different devices span widely differing ranges of operating conditions (geometry, magnetic field, temperature, ...). However, by focussing on those features that are common to all
experiments, one may hope of disentangling the physical mechanisms driving the turbulence from the 
contingent details, related to specific geometrical and/or operational features of each machine. \\ 
One purpose of the present paper is to present an extensive comparison of edge turbulence data taken from machines
in different configurations: two Tokamaks (NSTX \cite{nstx} and Alcator C-Mod \cite{cmod}) and two Reversed Field Pinches (TPE-RX \cite{tpe} and RFX-mod \cite{rfx}). 
The data were taken by our group using the same technique: Gas Puffing Imaging (GPI). This allows a straightforward intermachine comparison.\\
The most informative way of expressing experimental time series is in terms of their 
probability density function (PDF). PDFs from all devices do indeed share some broad features: 
by considering density fluctuation PDFs, they are strongly skewed curves
with approximately exponential tails towards high values, definitely departing from the Gaussian paradigm
of independently distributed small fluctuations. When dealing with large databases, however, using the whole PDF
becomes cumbersome, and more compact ways of expressing information must be found. 
Quite recently, Labit \textit{et al} \cite{labit} (hereafter referred to as Labit for short), 
in an analysis of TORPEX \cite{torpex} data, proposed of expressing data in terms of their
normalized third (Skewness $S$) and fourth statistical moment (Flatness $F$). Besides the convenience of replacing
a whole time series made of thousands of figures with a single couple of numbers, it was there argued that
$S$ and $F$ do actually contain a large part of the physics embedded in the raw data in terms of their mutual relation: 
indeed, empirical data from TORPEX as well as from many other physical systems  \cite{sura}, together with 
general analytical results, point to the fact that $S$ and $F$ are roughly related by a quadratic relation
$F = A\cdot S^2 + B\cdot S + C$, and the details about the physics driving the system are packed into the coefficients $A,B,C$ (although the linear term $B$ is usually discarded, either because it is empirically found
negligible, or because theoretical considerations suggest that it should be null. This issue will be more extensively discussed later).\\
In order to provide a physical rationale to the behaviour of a raw time series,
one can be content with just a phenomenological approach, i.e., picking up an analytical approximation 
to the empirical PDF, on the basis of the good agreement with data, but also backed up by some qualitative modelling of the physics running in background. 
This is the approach followed in the Labit paper
as well as in several earlier studies \cite{pdf,pop04,ppcf06,mast}. The coefficients $A,B,C$ may be compared 
with those directly computed from the guessed PDF. Alternatively, a first--principle approach may also be attempted, where some more fundamental equations are postulated to model the system studied. Time series may be generated by solving these equations, and the corresponding statistical moments evaluated. Fairly recently, Krommes \cite{krommes}, adapting to edge plasma turbulence 
an earlier model originally developed to study the fluctuations  of sea surface temperature \cite{sura} was able 
to reproduce a $F$--\textit{versus}--$S$ curve closely resembling TORPEX data \cite{labit}. It is particularly
remarkable that, although the relevant equations are stochastic nonlinear differential equations,
the computation of $S$ and $F$ in terms of the parameters of the model may be done in analytical closed form.   \\
In this work the data will be analyzed 
in terms of the moments $S,F$ and results will be compared with Labit'. 
We anticipate here that
some differences with those works will be found and discussed. The last part of the paper will
attempt to organize these novel features within a phenomenological model previously advanced \cite{ppcf06}.  

\section{The GPI diagnostics}
Before entering the discussion of the results, let us provide some more details about the instrumental
technique and data-taking procedure (more detailed reports are found in \cite{gpi1}). 
The GPI diagnostic is an optical, non-intrusive diagnostic for studying 
the plasma edge of fusion devices, and particularly the edge turbulence. It measures the visible 
light emission, usually emitted by neutral Hydrogen, Deuterium or Helium puffed
into the plasma edge. \\
In the two RFP experiments (RFX-mod and TPE-RX) the same GPI has been used, described
in ref.~\cite{Cavazzana04}. It consists of a gas-puffing nozzle, and an optical system for 
collecting the light emission. The optical diagnostic arrangement views edge regions in the plane 
perpendicular to the main component of the magnetic field, that is the radial-toroidal plane in the 
RFP experiments. It covers an observation region of about 80~mm along the toroidal direction 
and 40~mm in the radial one, by means of 32 lines of sight~(LoS). The signals are sampled at 10 Msamples/s. 
%\color{red}
GPI data are line-integrated measurements, although the effective length of integration
is limited by the volume where emission of the puffed neutral gas takes place, which is
very small (order a few cm). We note that this might a difference with respect to
electrostatic probes, which sample smaller regions of space. This issue will be the subject of 
further investigations.  
%\color{black}
The two-dimensional mapping
of the signal is obtained through a tomographic inversion \cite{ppcf07}. In this work, however, 
we will present just the signal from one chord. In order to be sure that conclusions do not depend upon the chord chosen, we made the following test: since we are interested in the third and fourth moment (skewness and flatness)
of timeseries, we computed them for all the RFX-mod pulses and for the seven central chords, and estimated
for each septet the mean value $<S>, <F>$, as well as the dispersion around this value. It turns out that
$95\%$ of points have a relative dispersion in $S$ lesser than $32\%$ and a dispersion in $F$ lesser than $18 \%$. 

In NSTX device the GPI consists of a double array of photomultiplier tubes that observes the
radial-poloidal plane, \textit{i.e.}, the plane perpendicular to the magnetic field at the edge of a
tokamak device (see references~\cite{Maqueda03,Zweben06}).
The signals are sampled at 0.5 Msamples/s. 
The GPI installed in Alcator C-Mod is quite similar, since it has 2 optical arrays that view the local gas puff in the plane perpendicular to the main magnetic field, and the light is collected by photodiodes \cite{terry}. The sampling frequency is 1 Msample/s.\\
The different arrangements imply that RFX-mod and TPE-RX measurements are line integrals along the radial
direction (with an effective width of order 2-3 cm), while NSTX and Alcator ones are along the toroidal direction. In principle, this could affect our results if measured quantities were strongly varying along the radial direction 
over distances smaller than about 1 cm.\\  
The whole database consists for RFX-mod of 2167 plasma discharges with different plasma
currents and densities; for TPE-RX, 40 plasma discharges at low plasma current (150-350~kA) have been considered;
for NSTX only 13 discharges with plasma current of about 800~kA and with a spontaneous L-H 
transition have been used. In this way, both the L- and the H-mode could be studied in the same
set of plasma shots. All the available GPI signals of NSTX have been considered, and so both
the poloidal and radial array of chords have been analyzed. The data 
are collected in different radial positions, from about 60~mm inside the separatrix up to about 60~mm outside.
Finally, we collected data from 10 Alcator discharges. 
Regarding typical plasma conditions, electron temperature in the collection volume is in the range 
10-40 eV for all devices, while typical plasma density is of order $10^{19}$ m${}^{-3}$ for RFX-mod, Alcator and
NSTX, and up to an order of magnitude smaller for TPE-RX.
All timeseries were splitted in strips 10 milliseconds long, and the couple $(S,F)$ was computed for each strip. Since the available part of the discharges is usually much longer than 10 ms, more than one $(S,F)$ couple was 
computed for each discharge. The time interval
was chosen as a good compromise between competing demands: it had to be as small as possible in order to 
consider as almost stationary the background plasma conditions during the single time series, but long enough to 
prevent undesired fluctuations due just to the scarce number of points, as well as spurious correlations due to 
microturbulence characteristic times, as guessed from power spectrum. Depending upon the device, hence, each couple 
$(S,F)$ was computed using up to $10^5$ points (for RFX-mod) down to 5000 (for NSTX). For comparison purposes, 
some trials were done, using RFX-mod data, by splitting time series into strips only 2 ms long. No relevant differerences were spotted on statistical terms: the average normalized difference $ < |S_{10ms} - S_{2ms}|/S_{10ms}>$ turns out to be less than $ 20 \%$ (the subscripts refer to moments computed using respectively the longer and the shorter time strips, and the brackets $<>$ imply an average over all pulses). 

\section{Intermachine comparison}
The main result of this section is summarized in the figures below, where we plot the database of couples $(S,F)$. 
Each subplot contains the database from one device. Overplotted, there is a subset of the RFX-mod database selected 
so as to keep matched a few operation parameters in both devices. 
This way, we attempted to compare data from roughly similar experimental conditions. 
In RFPs three meaningful parameters are: (I) the ratio $N/I_p$ between the (line-integrated) density and the plasma current, or equivalently $n/n_G$ between the average density $n$ and the Greenwald
density $n_G$ \cite{greenwald}; (II) the reversal parameter $F = B_{tor}(a)/<B_{tor}>$, ratio between the toroidal magnetic field at the edge and its value averaged over the whole plasma. In RFPs it is negative, since the toroidal field reverses sign close to the edge, and is used to parametrize the equilibrium. (III) Finally, the absolute value of plasma current $I_p$ is an important parameter, too. 
TPE-RX data (and the accompanying RFX-mod ones) are parametrized by: 
(I) low normalized density, $ 0.1 < n/n_G < 0.3$; (II) deep reversal parameter, $ F < - 0.08$;  
(III) low plasma current, $I_p$ lower than 350 kA. In tokamaks, the reversal parameter $F$ does not make sense
since it is trivially fixed by the aspect ratio. Hence,  
NSTX pulses are characterized by the normalized density: $ 0.25 < n/n_G < 0.45$ and the plasma current $I_p = 800 $ kA.
Alcator C-Mod pulses, finally, have $ 0.1 < n/n_G < 0.5 $, realized in correspondence of two values of $I_p = 0.4, 0.8$ MA (corresponding to two values of magnetic field: 2.4 and 5.4 T). In this case we did not overplot RFX data.

\begin{figure}
\includegraphics{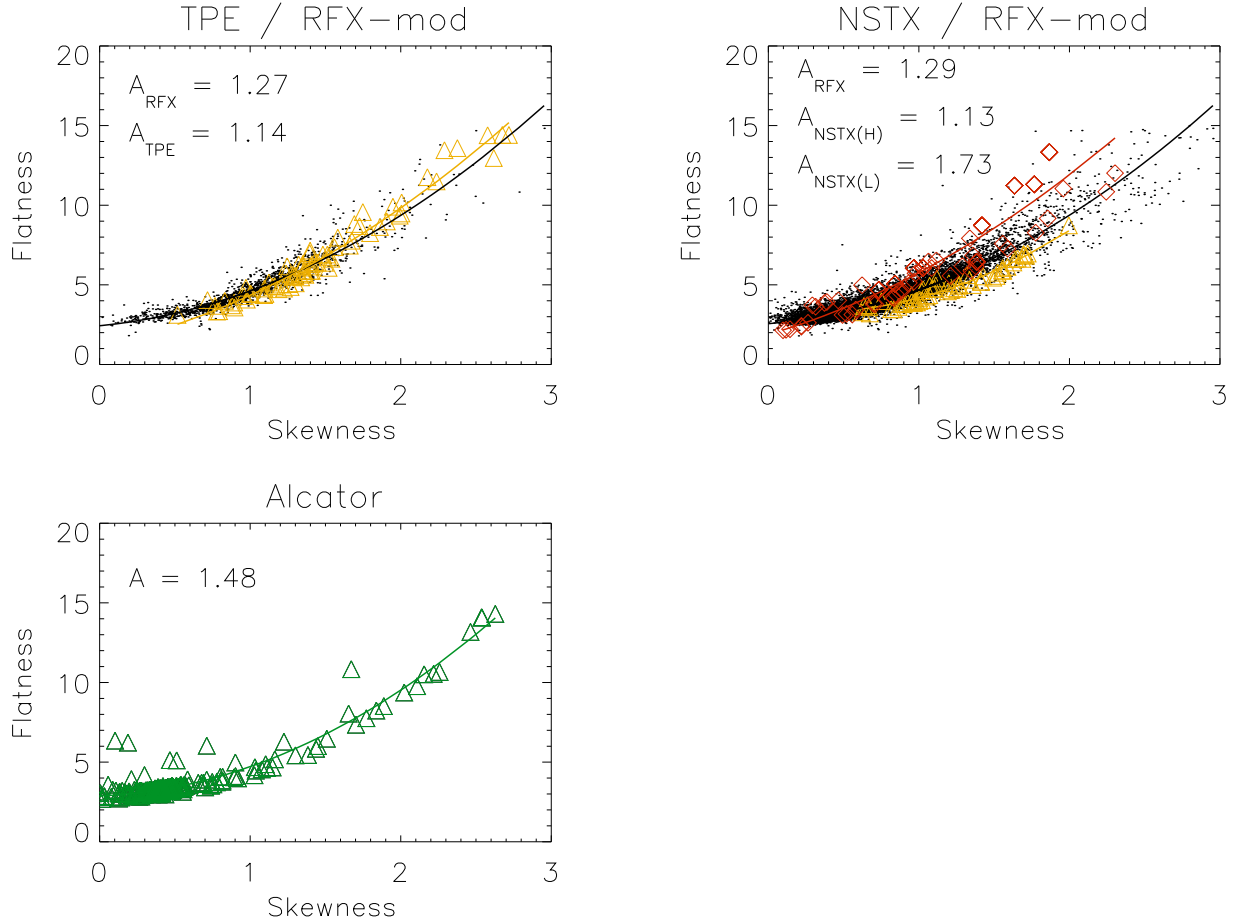} %{rfx_tpe_nstx_alc.eps}
\caption{(Color online) Top-left panel: yellow triangles, TPE data; dots, RFX-mod data selected so as to match
TPE parameters (see main text). Solid curves are interpolating second-degree polynomials 
$F = A\cdot S^2 + B\cdot S + C$, and the values of the coefficients $A$ are explicit displayed.
Top-right panel, the same for NSTX. Here, both the data in L-mode (yellow) and H-mode (red) are displayed.
RFX-mod data are selected to match NSTX parameters. Bottom panel, Alcator C-Mod data with interpolating polynomial.  }
\label{RFXTPENSTX}	
\end{figure} 

Several conclusions may be drawn from the plots (\ref{RFXTPENSTX}): data do indeed distribute around quadratic curves, as shown by Labit. However, there are sensible differences between experiments. 
Loosely speaking, results may be parametrized according to the degree of order that is commonly associated to each magnetic configuration: tokamaks data--with Alcator \textit{in primis}--align fairly well along a unique curve; NSTX in L-mode does the same, while in H-mode there is slightly more scattering  at large $S$. Interestingly, the L-mode and H-mode curves are definitely different. This is an hint
that the present analysis is sensitive to changes in the plasma transport. RFPs data are unequivocably scattered throughout a finite area. TPE-RX data appear more concentrated, but it could be
an artefact due to the poorer statistics. At this stage, we cannot still conclude whether the spread of the data 
around the interpolating curve is a consequence of the non-homogeneous set of plasma conditions, although
it appears a plausible conclusion. \\
In order to make more quantitative estimates, we have tabulated into tables I and II the numerical values 
of the fitting coefficients. In table I we reported the results for the most general fitting: 
$ F = A\times S^2 + B\times S + C$, while in Table II we did the same but postulating that the linear term
were not present, just like in Labit: $ F = A'\times S^2 + C'$. 
\begin{table}[ht]
	\centering
		\begin{tabular}{|c|c|c|c|}
\hline		experiment &  $A$ & $B$ & $C$  \\
\hline		 RFX-mod (plot a) & $ 1.27 \pm 0.05$ & $0.91 \pm 0.11 $ & $ 2.44 \pm 0.06$  \\
\hline		 TPE-RX & $1.14 \pm 0.20$ & $2.03 \pm 0.66$ & $ 1.17 \pm 0.52 $  \\
\hline     RFX-mod (plot b) & $ 1.29 \pm 0.01 $ & $ 0.82 \pm 0.03 $ & $2.57 \pm 0.01 $ \\
\hline     NSTX (L-mode) & $ 1.74 \pm 0.31 $ & $ -0.98 \pm 0.77$ & $3.42 \pm 0.46$ \\
\hline     NSTX (H-mode) & $ 1.13 \pm 0.40$ & $ 2.75 \pm 0.93 $ & $ 1.89 \pm 0.48 $  \\
\hline		 Alcator & $1.48 \pm 0.04 $ & $0.35 \pm 0.07 $ & $ 2.88 \pm 0.03 $\\
\hline			
		\end{tabular}
	\caption{Fitting coefficients for the full parabolic fit $F = A\times S^2 + B\times S + C$. 
	The rows RFX-mod (plot a) and RFX-mod (plot b) refer to the two subsets of RFX-mod data
	compared respectively to TPE-RX and NSTX data.}
	\label{tab:uno}
\end{table}

\begin{table}[ht]
	\centering
		\begin{tabular}{|c|c|c|}
\hline		experiment &  $A'$ & $C'$  \\
\hline		 RFX-mod (plot a) & $ 1.65 \pm 0.02 $ & $ 2.90 \pm 0.02 $   \\
\hline		 TPE-RX & $ 1.75 \pm 0.04 $ & $ 2.73 \pm 0.11 $\\
\hline     RFX-mod (plot b) & $1.67 \pm 0.005$ & $ 2.92 \pm 0.005$  \\
\hline     NSTX (L-mode) & $ 1.35 \pm 0.04 $ & $ 2.85 \pm 0.08 $ \\
\hline     NSTX (H-mode) & $ 2.28 \pm 0.10$ & $ 3.19 \pm 0.20 $  \\
\hline     Alcator & $ 1.62 \pm 0.02 $ & $ 2.97 \pm 0.02 $ \\
\hline			
		\end{tabular}
	\caption{Fitting coefficients just like in table I, for the reduced parabolic fit $F = A'\times S^2  + C' $.}
	\label{tab:due}
\end{table}
Obviously, the three-parameter fitting performs better in the least-squares sense than the two-parameters one.
However, it is interesting to notice that even the three-parameters fitting is not able to catch the whole physics
embedded in the data: if data do really align along a unique parabolic curve, then the residuals would be of purely
statistical origin, and therefore would be distributed according to a normal distribution. 
%\color{red}
D'agostino-Pearson's is a statistical test designed to assess whether a sample of data derives from
a normal distribution \cite{pearson}. We carried on such a test on our data and found that it yielded a negative answer (to within the standard 0.05 significance level): the scatter of data around the quadratic fitting curve is not distributed as expected from a normal distribution. %\color{black}
We note that some degree of scattering is present also in TORPEX data \cite{labit} but was not given
a particular emphasis in that analysis. We recall that we are not dealing here with raw data but with moments,
i.e., averages of the data. This means that fluctuations of purely statistical origin should-to a large extent-be washed out by the averaging procedure: it is easy to show that the sampling errors on $S, F$ due to the finite length of the time series are orders of magnitude smaller than the observed scatter. Therefore, there must be another cause
for the spreading of $(S,F)$, not related to insufficient statistics. An obvious guess is relating
$S,F$ to microscopic plasma conditions that are fluctuating from one measurement to another. \\
Also remarkable is that the coefficient
of the quadratic term is always close to, but usually significantly different from, the value $A = 1.5$ recovered for TORPEX data. The value $A=3/2$ was first justified in Labit arguing that the functional form of edge density fluctuations were closely matched by a Beta distribution \cite{beta}: an analytical curve often encountered 
in modelling statistics problems.   
The parabolic relation between $S$ and $F$, by itself, is not surprising; 
it is often reported in experiments from different environments (from atmospheric sciences to oceanography \cite{sura}).
Indeed, speculations may be put forth, supporting the view that it must be a fairly general result. Very recently, 
Krommes \cite{krommes} has developed a model for studying edge density fluctuations. An exact result of him is that 
$ F = (3/2 - a_0) S^2 + C(a_0)$, with $a_0$ a free parameter within the model. \\
Within Krommes' modelization, a change of $F$--\textit{versus}--$S$ trend may be taken into account 
through the coefficient $a_0$, that potentially may vary between any couple of measurements. On the contrary,
Labit analysis is rigid in prescribing $A = 3/2$. 
We further note that the interpolating curves feature non-negligible linear terms
$B\cdot S$. This is particularly evident for NSTX in H-mode, but is generically true for all the fits. Again, this result clashes with Labit' modelling 
using Beta distributions, but in this case also Krommes' model fails, since no linear term is there expected.  \\
NSTX fits feature striking differences between L-mode and H-mode, suggesting deep changes to the underlying physics. 
\begin{table}[ht]
	\centering
		\begin{tabular}{|c|c|c|c|c|c|c|}
\hline		experiment &  $A'$ & $C'$ & $\Delta'$ & $B''$ & $ C''$ & $\Delta''$ \\
\hline     NSTX (L-mode) & $ 1.35 \pm 0.04 $ & $ 2.85 \pm 0.08 $ & 4.44 & $ 3.31 \pm 0.13$ & $0.97 \pm 0.16$ & 6.72\\
\hline     NSTX (H-mode) & $ 2.28 \pm 0.10$ & $ 3.19 \pm 0.20 $  & 106 & $ 5.28 \pm 0.25$ & $ 3.19 \pm 0.21 $ & 105\\
\hline			
		\end{tabular}
	\caption{Fitting coefficients for the reduced parabolic fits $F = A'\times S^2  + C' $ (second and third column)
	and $ F = B''\times S + C''$ (fifth and sixth column). $\Delta'$ and $ \Delta''$ are the sum of residuals squared
	($\Delta = \sum_i (d_i - d_{i,fit})^2$) for the two fits respectively.}
	\label{tab:tre}
\end{table}
Table III points that, in H-mode, the $(S,F)$ is acquiring a stronger and stronger dependence from the linear term
at the expenses of the quadratic one: the linear fit perform as well as the quadratic one. \\
A further interesting issue is that of the existence of time series with negative skewness. In our measurements
they are very unlikely, and furthermore are rather marginal: $ - 1 < S $ (see Fig. 2).
Physically, negative skewness in density data implies that fluctuations are favoured depleting particles from 
the region of the measurements. The possibility of their existence was documented by DIII-D \cite{boedo},
TORPEX \cite{labit}, and Large Plasma Device \cite{carter}, that interpreted them as density holes. 
The exact reason of this discrepancy between our results and theirs is as yet unclear. We note that all of the other mentioned analysis have been carried on using edge Langmuir probes. The issue of the correspondence between quantities measured by probes and those measured by GPI is not still completely settled. 
Up to now, all investigations done by our group have not yet found anything invalidating
the view that  essentially the same physics is monitored by both diagnostics \cite{mago}. 
On the other hand, quite recently 
theoretical speculations have been put forth, suggesting that in some situations definite differences 
may be expected \cite{myra}.

The above results suggest therefore that a fully satisfactory modelling of edge 
fluctuations, valid for all devices, has not yet been achieved. In the next section we reconsider 
an instance of such a modelization, advanced in an earlier work, and show to what extent it may fit the present 
results. 
   
\section{An attempt of phenomenological modelling of edge fluctuations} 
\begin{figure}
\includegraphics[width=90mm]{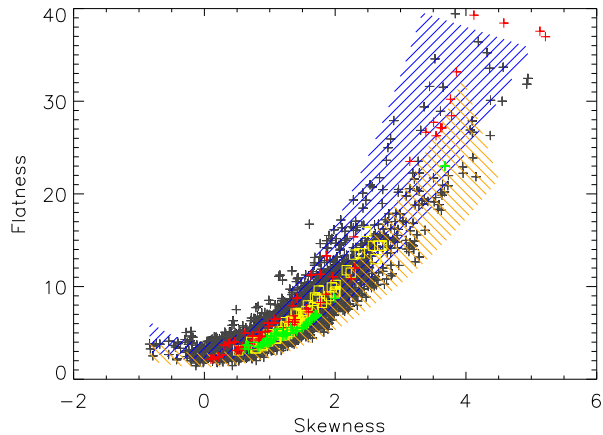} %{fig1.eps}
\caption{(Color online) Symbols, GPI data from different experiments: grey crosses, from RFX-mod; yellow squares, TPE-RX;
red crossed, NSTX in H-mode; green crosses, NSTX in L-mode. The blue region fills 
the area spanned by the two-Gamma PDFs; orange region is the two-Beta PDFs. }
\label{figura1}	
\end{figure}

A first attempt of phenomenological modelling of GPI data in RFX-mod was given in \cite{ppcf06}. The need for 
a different modelling tool with respect to existing proposals was stimulated by the recognition in 
RFX-mod data of a novel feature, unnoticed in other experiments, i.e., the frequent presence of a double slope in the tail of the PDF \cite{ppcf06}. 
We may now confirm that a similar feature is spotted also in other devices (Alcator C-Mod and TPE-RX). 
In \cite{ppcf06} we justified this double slope
 by postulating the existence of two contributions to the total signal, hence that total PDF be a linear 
 combination of two basis functions:
\begin{equation}
	P(x) = c_1 P_1(x) + c_2 P_2(x)
	\label{duepdf}
\end{equation}
In the original work \cite{ppcf06} we used Gamma distributions as basis functions:
\begin{equation}
	P_j(x) = {(\beta_j N_j)^{N_j} \over \Gamma(N_j)} x^{N_j -1} \exp\left(-\beta_j N_j x\right), \quad 
	         \Gamma(z) = \int_0^\infty t^{z-1} \exp(-t)\, dt
	\label{gamma}         
\end{equation}
Gamma distributions are fairly versatile, and may fit a large variety of empirical curves.
They satisfy an important constraint: are defined only for 
$x > 0$, hence prevent density being negative. This was somewhat overlooked in earlier studies, where 
fluctuations around the average value $ x - <x>$ were considered, rather than the whole signal, missing
therefore an important information. \\
The parameter $N_j$ plays the role of a number of degrees of freedom: it can be shown 
that $P_j$ is the resulting distribution when our signal is made by the sum of $N_j$ independent stochastic 
variables $y^{(j)}_l , l = 1,...,N_j$, each of them being distributed according to a Boltzmann distribution:
\begin{equation}
	P(y)= {1 \over <y>} \exp\left(-{ y \over <y>}\right)
	\label{boltzmann}
\end{equation}
Since $N_1 \neq N_2$, the two contributions in (\ref{duepdf}) 
were thought as coming from two mechanisms living in abstract spaces of different dimensionality. 
When $N \to \infty$, Gamma distributions approach Gaussians, hence it is natural to relate the high-$N$ contribution to turbulence coming from uncorrelated small scale fluctuations, while the low-$N$ one is modelling some correlation
that diminishes the effective number of degrees of freedom. See \cite{farge} for an earlier suggestion pointing to this kind of mechanism. Note that Eq. (\ref{duepdf}) refers to addition of probabilities, not of densities. In other terms, we are postulating that, at each time step, the signal is due \textit{entirely} either to one mechanism or to the other, but not to both. This is clearly just a first approximation.  \\
It is obvious that the choice of the basis function is constrained only by the final accordance with 
experiment, hence we may inspect the effect of using different basis function. 
Accordingly, we thought of straightforwardly generalizing TORPEX analysis, using as basis functions $P_1, P_2$ \textit{two} Beta distributions:
\begin{eqnarray}
P_j(x) \equiv P(x,p_j,q_j,x_j^{(l)},x_j^{(h)}) &=& { (x-x_j^{(l)})^{p_j - 1} (x_j^{(h)} - x)^{q_j -1} \over 
                                           {\cal B}(p_j, q_j) (x_j^{(h)} - x_j^{(l)})^{p_j + q_j -1}} \nonumber \\
                                           \quad {\cal B}(p,q) &=& \int_0^1 t^{p -1} (1 - t)^{q - 1} dt 
\label{beta}
\end{eqnarray}
[Note that we use the symbols $\beta_j$ within the
definition of Gamma PDFs (\ref{gamma}). This turns out a bit unfortunate in the present context because of
possible confusion with Beta distributions, but is needed in order to be consistent with earlier literature]. \\
From Eq. (\ref{duepdf}) we then evaluate the mean value and higher moments:
\begin{eqnarray}
<x> &=& \int x P(x) dx \nonumber \\
M(i) &=& \int (x - <x>)^i P(x) dx , i > 1 
\label{momenti}
\end{eqnarray} 
in particular, $S = M(3)/M(2)^{3/2}$, $F = M(4)/M(2)^2$. 
Although explicit analytical formulas may be given for $S$ and $F$ when $P_j$ are given by Eqns. (\ref{gamma}) or (\ref{beta}), we won't write them down here, since 
they are fairly lengthy. 
It is however obvious that expressions (\ref{momenti}) depend on a large number of free parameters:
there are six of them in the case of two Gamma distributions, but two are fixed by requiring normalization of $P$: 
$c_1 + c_2 = 1$, and of average density: $c_1/\beta_1 + c_2/\beta_2 = 1$.    
For Beta distributions, the situation is more intricate, since we start with a larger number of parameters: ten. 
We still get rid of one of them by the normalization condition on $P$. An interesting property of one \textit{single}
Beta distribution is that $S, F$, do not depend upon the boundaries $x^{(l)} , x^{(h)}$.  In the two-PDF case,
this result does not hold any longer because of the way the dependences upon the parameters 
are hardwired into $<x>$. In order to simplify algebraic expressions, we considered $<x>$ as a supplementary free parameter. Since it is formally independent of $\{x_j^{(l)}, x_j^{(h)}\}$, these latter quantities could therefore be discarded from final results. We are thus left with six free parameters. In principle, therefore, Beta distributions have more room for interpolation.        \\
Figure (\ref{figura1}) summarizes the result of the analysis. Symbols are all 
the measurements from the three devices RFX-mod, TPE-RX and NSTX \cite{nota}, the shaded
regions fill the parameter space spanned by the analytical curves (\ref{duepdf})
through a scan of the free parameters over their domains.  It is interesting to notice that, 
although \textit{one} Gamma PDF cannot have negative skewness, 
a linear combination of \textit{two} Gamma PDFs can--at least marginally.
Notwithstanding the fact that fitting with Beta PDFs ought be 
easier, since we have more free parameters available, actually these functions turn out to be less flexible: 
there is not a large difference between using just one or a sum of two Beta PDFs. \\
On the whole, the blue region overlaps better the cloud of experimental data: most of them fall 
inside it. On the contrary, about one half of the points lies outside the orange
region. Hence, we can claim that using a linear combination of two Gamma PDFs should, on the average, 
perform better on our data.  However, the purpose of our work was not as much assessing which 
analytical PDF works better (in principle, there could be other better choices until now unexplored)
as establishing the minimal degree of sophistication needed for a good modelling of edge fluctuations.  

\section{Summary and conclusions}
Summarizing, we have reached the following conclusions: 
(I) The physically relevant issue of the \textit{universality} of edge turbulence between 
different devices has been addressed, with mixed results. From the one hand, we may confirm earlier studies:
all data align along parabolic curves. However, we pointed out that the mere existence of a quadratic relation between
$S$ and $F$ does not provide much insight. The relevant information is contained in the coefficients
of the curve, that appear not univocal: they are very similar between two RFPs experiments
(TPE-RX and RFX-mod), differ between a tokamak and a RFP (NSTX and RFX-mod), as well as between 
two modes of operation of the same tokamak (NSTX in L-mode and H-mode), and, finally between 
two tokamaks (NSTX and Alcator C-Mod). Furthermore, within one and the same machine, the spread
of points in the $(S,F)$ plane warns that plasma parameters varying within the same shot strongly
affect the statistics of the turbulence.
(II) The analysis in terms of $(S,F)$ introduced in Labit confirms to be a very effective
and compact tool.    
(III) As a consequence of (I), any attempt of modelling phenomenologically the 
statistics of edge turbulence must be fairly flexible. The model advanced in the previous section
appears promising under this respect.

\begin{acknowledgments}
This work was supported by the European Communities under
the contract of Association between EURATOM/ENEA. The views and opinions expressed herein do not necessarily reflect those of the European Commission. The TPE-RX program was financially supported
by the Budget for Nuclear Research of the Ministry of Education, Culture, Sports, Science
and Technology, based on the screening and counselling of the Atomic Energy Commission. S. Cappello read the
manuscript and provided several useful suggestions.
\end{acknowledgments}


\begin{thebibliography}{99}

% [1]
\bibitem{modelli} O.E. Garcia, J. Horacek, R.A. Pitts, A.H. Nielsen, 
W. Fundamenski, J.P. Graves, V. Naulin and J. Juul Rasmussen, 
Plasma Phys. Control. Fusion \textbf{48}, L1 (2006);
N. Bisai, A. Das, S. Deshpande, R. Jha, P. Kaw, A. Sen, and R. Singh,
Phys. Plasmas \textbf{12}, 072520 (2005);
B. Scott, Phys. Plasmas \textbf{12}, 062314 (2005);
S. Servidio, \textit{et al}, Phys. Plasmas \textbf{15}, 012301 (2008);
G.N. Kervalishvili, \textit{et al}, Contrib. Plasma Phys. \textbf{48}, 32 (2008)

% [2]
\bibitem{soc} D.E. Newman, B.A. Carreras, P.H. Diamond, T.S. Hahm, Phys. Plasmas \textbf{3}, 1858 (1996);
B.A. Carreras, D. Newman, V.E. Lynch, P.H. Diamond, Plasma Phys. Rep. \textbf{22}, 740 (1996);
B.A. Carreras, \textit{et al} Phys. Rev. Lett. \textbf{80}, 4438 (1998)

% [3]
\bibitem{pdf} 
G.Y. Antar, S.I. Krasheninnikov, P. Devynck, R.P. Doerner, E. M. Hollmann, 
J.A. Boedo, S.C. Luckhardt, and R.W. Conn, Phys. Rev. Lett. \textbf{87}, 065001 (2001);
G.Y. Antar, P. Devynck, X. Garbet, S. C. Luckhardt, Phys. Plasmas \textbf{8}, 1612 (2001);
G.Y. Antar, G. Counsell, Y. Yu, B. Labombard, P. Devynck, Phys. Plasmas \textbf{10}, 419 (2003);
J.P. Graves, J. Horacek, R.A. Pitts and K.I. Hopcraft,
Plasma Phys. Control. Fusion \textbf{47}, L1 (2005);
F. Sattin, N. Vianello, Phys. Rev. E \textbf{72}, 016407 (2005)

% [4]
\bibitem{pop04} F. Sattin, N. Vianello, M. Valisa, Phys. Plasmas \textbf{11}, 5032 (2004)

% [5]
\bibitem{ppcf06} F. Sattin, P. Scarin, M. Agostini, R. Cavazzana, G. Serianni, M. Spolaore and N. Vianello,
Plasma Phys. Control. Fusion \textbf{48}, 1033 (2006)

% [6]
\bibitem{labit}  
B. Labit, I. Furno, A. Fasoli, A. Diallo, S. H. Müller, G. Plyushchev, M. Podest$\acute{a}$, and F. M. Poli, 
 Phys. Rev. Lett. \textbf{98}, 255002 (2007)


% [7] 
\bibitem{mast} M. $\check{S}$kori$\acute{c}$ and M. Rajkovi$\acute{c}$, 
    Contrib. Plasma Phys. \textbf{48}, 37 (2008)
 
% [8] 
\bibitem{pedrosa} M.A. Pedrosa, \textit{et al}, Phys. Rev. Lett. \textbf{82}, 3621 (1999); 
E. Sanchez, \textit{et al}, Phys. Plasmas \textbf{7}, 1408 (2000) 

% [10]
\bibitem{nstx} M.Ono, M.G.Bell, R.E. Bell, T. Bigelow, \textit{et al.}, Plasma Phys. Control. Fusion \textbf{45}, A335 (2003)

\bibitem{cmod} I.H. Hutchinson \textit{et al} Phys. Plasmas \textbf{1}, 1511 (1994)

% [12]
\bibitem{tpe} Y. Yagi, S. Sekine, T. Shimada, A. Masiello, \textit{et al.}, Fusion Eng. Design \textbf{45}, 421 (1999)

% [13]
\bibitem{rfx} S. Martini, M. Agostini, C. Alessi, A. Alfier, \textit{et al}, Nucl. Fusion \textbf{47}, 783 (2007)

%[14]
\bibitem{torpex} A. Fasoli, B. Labit, M. McGrath, S. H. Müller, G. Plyushchev, M. Podest\'a, and F. M. Poli, 
Phys. Plasmas \textbf{13}, 055902 (2006)

\bibitem{sura} A. Maurizi, Nonlin. Processes Geophys. \textbf{13}, 119 (2006);
               S. Alberghi, A. Maurizi, F. Tampieri, Journal of Applied Meteorology \textbf{41}, 885 (2002);
               T. P. Schopflocher and P. J. Sullivan, Boundary-Layer Meteorology \textbf{115}, 341 (2005);
               P. Sura and P. D. Sardeshmukh, J. Phys. Oceanogr. \textbf{38}, 638 (2008);
               K.R. Sreenivasan and R.A. Antonia, Annu. Rev. Fluid Mech. \textbf{29}, 435 (1997);
               A. Bertelrud, S. Johnson, C. Lytle, C. Mills, 
in Proceedings of the International Congress on Instrumentation in Aerospace Simulation Facilities 
(1997). Full text available at the URL http://ieeexplore.ieee.org.   
%``A system for analysis of transition characteristics on a high-lift configuration
%at high Reynolds numbers'', in Proceedings of 
%the International Congress on Instrumentation in Aerospace Simulation Facilities 
%(1997). Full text available at the URL http://ieeexplore.ieee.org.   


\bibitem{krommes} J.A. Krommes, Phys. Plasmas \textbf{15}, 030703 (2008)

% [15]
%\bibitem{sura} A. Maurizi, Nonlin. Processes Geophys. \textbf{13}, 119 (2006);
%P. Sura and P. D. Sardeshmukh, Journal of Physical Oceanography \textbf{38}, 638 (2008)

%\bibitem{horton}W. Horton and Y.H. Ichikawa, \textit{Chaos and Structures in Nonlinear Plasmas}
%(World Scientific, 1996), ch. 6.7

% [9]
\bibitem{gpi1} M. Agostini, R. Cavazzana, P. Scarin, G. Serianni, Rev. Sci. Instrum. \textbf{77}, 10E513 (2006);
P. Scarin, M. Agostini, R. Cavazzana, F. Sattin, G. Serianni, N. Vianello, Jour. Nucl. Mat. \textbf{363}, 669 (2007); 
R. Cavazzana, G. Serianni, P. Scarin, M. Agostini, \textit{et al}, Plasma Phys. Control. Fusion \textbf{49}, 129 (2007)

% [20]
\bibitem{Cavazzana04} R. Cavazzana, P. Scarin, G. Serianni, M. Agostini, \textit{et al.}, Rev. Sci. Instrum.  \textbf{75}, 4152 (2004)

% [21]
\bibitem{ppcf07} G. Serianni, M. Agostini, R. Cavazzana, P. Scarin, Plasma Phys. Contr. Fusion \textbf{49}, 2075 (2007)

% [22]
\bibitem{Maqueda03} R.J. Maqueda, G.A. Wurden, D.P. Stotler, S.J. Zweben, \textit{et al},
Rev. Sci. Instrum.  \textbf{74}, 2020 (2003)

% [23]
\bibitem{Zweben06} 
S.J. Zweben, R.J. Maqueda, J.L. Terry, T. Munsat, \textit{et al.}, Phys. Plasmas,  \textbf{13}, 056114 (2006)

\bibitem{terry}
J.L. Terry \textit{et al.}, Phys.Plasmas \textbf{10}, 1739 (2003); 
J.L. Terry \textit{et al.}, Jour. Nucl. Mat. \textbf{363-365}, 994 (2007)

% [26]
\bibitem{boedo} J.A. Boedo, \textit{et al}, Phys. Plasmas \textbf{10}, 1670 (2003) 

% [27]
%\bibitem{gpi2} M. Agostini, S.J. Zweben, R. Cavazzana, P. Scarin, G. Serianni, R.J. Maqueda, D.P. Stotler, Phys. Plasmas \textbf{14}, 102305 (2007)

\bibitem{greenwald} M. Greenwald, Plasma Phys. Control. Fusion \textbf{44}, R27 (2002)


% [17]
%\color{red}
\bibitem{beta} M. Evans, N. Hastings, B. Peacock, \textit{Statistical Distributions}, (Wiley, 2000)
%\color{black}

% [25]
\bibitem{carter} T.A. Carter, Phys. Plasmas \textbf{13}, 010701 (2006)

\bibitem{mago} M. Agostini, et al., Plasma Phys. Control. Fusion \textbf{50}, 095004 (2008)

\bibitem{myra} D.A. Russell, et al, Bull. Am. Phys. Soc. \textbf{53}, 193 (2008)

\bibitem{farge} M. Farge, K. Schneider, P. Devynck, Phys. Plasmas \textbf{13}, 042304 (2006)

\bibitem{nota} Alcator data became available after this part of analysis had been carried out. Although
they could easily be inserted in the figure, we chose not to do it in order not to clog the plot with
too many details without adding substantial information.

%\color{red}
\bibitem{pearson} D.J. Sheskin, \textit{Handbook of Parametric and Nonparametric Statistical Procedures},
(Chapman and Hall/CRC, 2000)
%\color{black}


% [16]
%\bibitem{sfteorico} F. Sattin \textit{et al}, in preparation


% [18]
%\bibitem{flynn} M.R. Flynn, Annals of Occupational Hygiene \textbf{48}, 491 (2004)

% [19]
%\bibitem{springer} See under the voice ``Beta distribution'' at Springer's online Encyclopaedia of Mathematics,
%URL: http://eom.springer.de

% [24]
%\bibitem{nicola} N. Vianello, technical note (unpublished)

% [2]
%\bibitem{serianni07} G. Serianni, \textit{et al}, Plasma Phys. Control. Fusion \textbf{49}, B267 (2007)



\end{thebibliography}
\end{document}